\documentclass[11pt]{article}
\pdfoutput=1

\usepackage{jheppub}
\usepackage[latin1]{inputenc}
\usepackage{amsmath}
\usepackage{amsfonts}
\usepackage{amssymb,MnSymbol,slashed}
\usepackage{amsxtra}
\usepackage{color}
\usepackage{hyperref}
\hypersetup{linktocpage=true}
\usepackage{graphicx}
\usepackage{placeins}
\usepackage{upgreek} 
\usepackage{mathrsfs}
\usepackage{accents}
\usepackage{multirow} 
\usepackage{tikz}
\usepackage{calc}
\usepackage{multicol}

\usepackage{color}
\definecolor{darkgreen}{rgb}{0,0.5,0}
\definecolor{darkblue}{rgb}{0,0,0.7}
\definecolor{darkred}{rgb}{0.5,0,0.0}
\definecolor{darkorange}{rgb}{0.8,0.4,0.0}

\newcommand{\beq}{\begin{equation}}
\newcommand{\eeq}{\end{equation}}
\def\be {\begin{equation}}
\def\ee {\end{equation}}
\def\bs#1\es{\begin{split}#1\end{split}}
\def\ba#1\ea{\begin{align}#1\end{align}}
\def\baed#1\eaed{\begin{aligned}#1\end{aligned}}
\def\bged#1\eged{\begin{gathered}#1\end{gathered}}
\def\bea{\begin{eqnarray}}
\def\eea{\end{eqnarray}}

\newcommand{\bseq}{\begin{subequations}}
\newcommand{\eseq}{\end{subequations}}

\def\nn{\nonumber}

\def\c{\chi}

\def\e{\epsilon}

\def\Im{\text{Im}}

\let\foo\bar 
\renewcommand{\bar}[1]{ {\foo{  #1} }{} }

\newlength{\dhatheight}

\def\Ftau{F(U, \bar{U})}
\def\Frho{F(T, \bar{T})}

\newcommand{\T}{T=B_{89}+i V_{T^2}}
\newcommand{\U}{U=U_1 + i U_2}
\newcommand{\leou}{ln\left(\frac{\eta^{24}(U)}{\bar{\eta}^{24}(\bar{U})} \right) }
\newcommand{\leot}{ln\left(\frac{\eta^{24}(T)}{\bar{\eta}^{24}(\bar{T})} \right) }

\newcommand{\sz}{SL(2,\mathbb{Z})}
\newcommand{\sr}{SL(2,\mathbb{R})}

\newcommand{\ett}{E_8 \times E_8}
\newcommand{\son}{\frac{SO(2,n)}{SO(2)\times SO(n)}}

\title{Discrete anomalies in supergravity and consistency of string backgrounds}

\author[a]{Ruben Minasian}
\author[a]{Soumya Sasmal}
\author[b]{and Raffaele Savelli}

\preprint{\begin{tabular}{r} IPhT-T16/156 \\  IFT-UAM/CSIC-16-133\end{tabular}}

\affiliation[a]{Institut de Physique Th{\'e}orique, Universit{\'e} Paris Saclay, CEA, CNRS,\\ Orme des Merisiers, F-91191 Gif-sur-Yvette, France}

\affiliation[b]{Instituto de F\'isica T\'eorica UAM-CSIC, Cantoblanco, 28049 Madrid, Spain}

\emailAdd{ruben.minasian AT cea.fr}
\emailAdd{soumya.sasmal AT u-psud.fr}
\emailAdd{raffaele.savelli AT uam.es}

\abstract{We examine $SL(2, \mathbb{Z})$ anomalies in ten and eight-dimensional supergravities, the induced local counterterms and their realization in string theory. Composite connections play an important r\^ole in the cancellation mechanism. At the same time their global properties lead to novel non-trivial consistency constraints on compactifications.}

\begin{document}

\maketitle

\section{Introduction}\label{sec:intro}

The moduli space of supergravity theories with high enough supersymmetry has the structure of a coset manifold (or of a product thereof), typically denoted by $G/H$. The numerator denotes the U-duality group of the theory, whose discrete version $G_\mathbb{Z}$ gives rise to an exact symmetry (in fact, a superselection rule) after quantization, whereas the denominator (the maximal compact subgroup of $G$) is regarded as a gauge symmetry of the theory. In particular, the supersymmetry variations of all fermionic fields, which are inert under $G$,  involve the gauge composite connection corresponding to $H$.
The content of physical particles of the theory is usually identified by fixing the gauge, thereby eliminating the redundant bosonic degrees of freedom associated to $H$. In some cases, however, $H$ contains a $U(1)$ factor, which may couple to fermions in a chiral fashion, a priori giving rise to a chiral anomaly.

When the local symmetry is gauge fixed, the U-duality becomes  non-linearly realized. Moreover, the fermionic fields now transform under $G$. This transformation may still be realized as a phase shift. Therefore, the gauge fixing translates the $U(1)$ anomaly into a $G_\mathbb{Z}$ one, making the theory ill-defined \cite{Marcus:1985yy}. This anomaly can be canceled by the addition of a local counterterm with appropriate $G_\mathbb{Z}$-modular properties. 

The easiest instance of such a phenomenon appears in maximal eight-dimensional (8D) supergravity. The relevant part of the coset is given by $SL(2, \mathbb{R})/U(1)$. The ensuing local counterterm involves a modular function in the complex scalar parametrizing the coset and provides a higher-derivative ($\sim \alpha'^3$) correction to the action. Similar behaviour is displayed by 8D theories with 16 supercharges which have a more complicated coset structure.

The pure ten-dimensional (10D) type IIB supergravity does not have an anomaly in spite of having $SL(2, \mathbb{R})/U(1)$ coset. However the presence of  seven-branes generates an $\sz$ anomaly  \cite{Gaberdiel:1998ui}. The fact that the anomaly is generated by seven-branes makes one wonder if the counterterm is related to higher-curvature couplings on the  seven-brane worldvolume.

In this paper we try to address the above question in the type IIB context. Moreover, we examine this type of anomalies in various supergravity theories in eight dimensions, with both maximal and minimal supersymmetry and with several choices of gauge group. The aim is to compare the higher-derivative terms fixed by anomaly cancelation to the effective couplings derived from the relevant string amplitudes. In some cases we find a perfect agreement of the higher-derivative structure. In 8D, this happens for the maximal supergravity and for the minimal one with $SO(32)$ and $\ett$ gauge groups. For other gauge groups in 8D there appear additional higher-derivative structures in the amplitude-induced effective couplings. For the cases of 8D non-Abelian symmetry of rank 16, we give an interpretation of such new structures in terms of massive states arising from the breaking of $SO(32)$ or of $\ett$. The details of such amplitude computations, crucial for this comparison, will appear separately in 
\cite{Sasmal:2016fap}, which fills some gaps in the existing literature on the one-loop five-point amplitudes.

Compactifications on K\"ahler manifolds of positive curvature and the role of composite connections is another  aspect explored in this paper. In fact, tadpole cancellation relates the curvature of the composite connection to the curvature of the compactification space. Hence the global properties of the former  impact the consistency of lower-dimensional theories. In particular, we note the importance of the massive states for the anomaly cancellation in six-dimensional (6D) theories obtained from an $S^2$ reduction of 8D minimal theories.

The outline of the paper is as follows. In section \ref{sec:green-gaberdiel}, we review the $\sz$ anomaly in 10D type IIB theory with seven-branes, and investigate its possible relation to brane couplings. Section \ref{sec:N=2} is devoted to the study of the  $\sz$ anomaly in  maximal 8D supergravity. Section \ref{sec:N=1}
does the same for minimally supersymmetric 8D theories with gauge groups $SO(32)$, $\ett$, $SO(16)^2$ and $SO(8)^4$, and contains a detailed comparison with string amplitude results.  Global constraints on the composite connections are discussed in section \ref{sec:glo}. The concluding section \ref{sec:DO} outlines some implications for the  consistency of compactifications to lower-dimensional theories. Conventions and some technical details are collected in three appendices.
    
\section{The $\sz$ anomaly in type IIB supergravity}\label{sec:green-gaberdiel}
A very interesting instance of composite $U(1)$ anomaly in supergravity theories arises already in the 10D type IIB theory \cite{Gaberdiel:1998ui}. The purpose of this section is to briefly review its cancellation mechanism, which will be common to all the anomalies discussed in this paper. Furthermore, we will speculate on some suggestive implications this particular 10D anomaly appears to have, especially concerning higher-derivative couplings on D7-brane worldvolumes.

\subsection{The Green-Gaberdiel counterterm}

The 10D type IIB theory has a global $\sr$ symmetry. The $SL(2,\mathbb{R})$ group manifold can be parametrized by a complex scalar $\tau = \tau_1 +i \tau_2$ (identified with the axio-dilaton field) taking values in the upper half plane, and a real (angular) scalar $0\le\phi\le2\pi$, which is a pure-gauge degree of freedom charged under the local symmetry group $U(1)\subset\sr$. The scalar manifold of the theory, i.e. the coset space $\sr/U(1)$, is then usually described by the following vielbein
\begin{equation}
\label{vielbein}
V^a_i=\frac{1}{\sqrt{-2i\tau_2}}\left( 
\begin{matrix}
\bar{\tau}e^{-i\phi} & \quad \tau e^{i\phi}\\e^{-i\phi} & \quad e^{i\phi} 
\end{matrix}\right) .
\end{equation}
Under a general $SL(2,\mathbb{R}) \times U(1)$ transformation, the vielbein $V^a_i$ transforms as 
\begin{equation}
\label{vtransf}
V'^a_i=A^a_b V^b_j u^j_i,
\end{equation}
where $A$ is the $\sr$ matrix 
\begin{equation}
A=\left(\begin{matrix}
a & \quad b\\
c & \quad d
\end{matrix} \right) , \quad a, b,c,d \in \mathbb{R}, \quad ad-bc=1,
\end{equation}
and $u$ is 
\begin{equation}
u=\left( \begin{matrix}
e^{-i\Sigma} & 0\\
0 & e^{i\Sigma}
\end{matrix}\right) \,, \qquad 0\le\Sigma\le2\pi\,,
\end{equation}
which thus shifts $\phi$ as $\phi\to\phi+\Sigma$.
The composite $U(1)$ connection is locally the $\sr$-invariant combination
\begin{equation}\label{CompConn}
Q_{\mu}= \partial_{\mu}\phi -\frac{\partial_{\mu}\tau_1}{2\tau_2},
\end{equation}
so that its $U(1)$ field strength takes the local form
\begin{equation}\label{CompFS}
F=dQ= \frac{d\tau \wedge d\bar{\tau}}{4i \tau_2^2}.
\end{equation}
The two gravitini form a complex conjugate pair which carry charges $\pm \frac{1}{2}$ under $U(1)$, and the two dilatini form a complex conjugate pair of opposite chirality and of charges $\pm \frac{3}{2}$. Due to these chiral couplings, the theory may suffer from an anomaly for the $U(1)$ gauge symmetry. At the perturbative level, this anomaly can be detected from one-loop hexagon diagrams containing at least one composite gauge field \eqref{CompConn}. Alternatively, it can be seen to descent from a 12-form anomaly polynomial, which, according to the rules summarized in appendix \ref{sec:index}, takes the form:
\begin{equation}\label{AnoPol}
P_{\rm 12}=\frac{F^2}{2\pi}\left[2X^-_8(R)+\frac{p_1(R)}{48}\left(\frac{F}{2\pi}\right)^2-\frac{1}{32}\left(\frac{F}{2\pi}\right)^4\right]\,,
\end{equation}
where we defined 
\begin{equation}\label{X8pm}
X^\pm_8(R)=\frac{1}{192(2\pi)^4}\left(trR^4 \pm \frac{1}{4}(trR^2)^2 \right)\,,
\end{equation}
in terms of the 10D Einstein-frame curvature two-form $R$, and $p_1(R)=-\tfrac12 trR^2/(2\pi)^2$ is the first Pontryagin class. The absence of an $F^0$-term in the expression \eqref{AnoPol} is clearly due to the well-known freedom of the type IIB theory from pure local gravitational anomalies. Moreover, the absence of a linear term in $F$ implies that the new $U(1)$ anomaly vanishes for a pure supergravity theory (i.e. without brane sources). Indeed, if no 7-brane is present, $F$ is an exact form and, because of its composite structure \eqref{CompFS}, it squares to zero. On the contrary, when 7-branes are there, the expression \eqref{CompFS} is only valid away from them, because the background value of $\tau$ undergoes monodromies around such sources.

From the anomaly polynomial \eqref{AnoPol} one deduces the anomalous phase variation of the partition function
\begin{equation}\label{VarPathInt}
 \Delta=-\int \Sigma\,\left[2 X_8^-(R)+\frac{p_1(R)}{48}\left(\frac{F}{2\pi}\right)^2-\frac{1}{32}\left(\frac{F}{2\pi}\right)^4\right]\, \frac{F}{2\pi}\,,
  \end{equation}
which can clearly be cancelled by the addition of the following counterterm in the 10D action:\footnote{Note that, contrary to \cite{Gaberdiel:1998ui}, higher powers of $F$ appear in $S_{\phi}$. The assertion of \cite{Gaberdiel:1998ui}  that $F^2=0$ is only true in the absence of 7-branes. Only in this case the expression \eqref{CompFS} is well defined. Otherwise $F^2$ is a form which localizes on the 7-brane worldvolume. As we will reiterate later, in the absence of 7-branes the anomaly is absent. }
 \begin{equation}
\label{GGu1}
  S_{\phi}=\int \phi\,\left[2 X_8^-(R)+\frac{p_1(R)}{48}\left(\frac{F}{2\pi}\right)^2-\frac{1}{32}\left(\frac{F}{2\pi}\right)^4\right]\, \frac{F}{2\pi}\,. 
  \end{equation}
 This formalism is notoriously redundant. However, upon fixing the gauge (say by setting $\phi\equiv0$),  symmetries will be realized non-linearly. Nevertheless, one can still describe the transformation properties of all fermion fields as local phase shifts, by specifying their charge under the $U(1)$ gauge symmetry. This is achieved simply by exploiting the property of the vielbein \eqref{vielbein} to convert $\sr$ indices into $U(1)$ indices. The result is that, in the gauge fixed theory, any field $\Psi$ with charge $q$ under the local $U(1)$ will have the following transformation under $\sr$:
\begin{equation}\label{GeneralPsiTransf}
\sr\ni \left(\begin{array}{cc}a&b\\c&d\end{array} \right)\,:\;\;\Psi\longrightarrow e^{iq\Sigma(\tau)}\Psi\qquad{\rm with}\qquad\Sigma(\tau)=-{\rm arg}(c\tau+d)\,.
\end{equation}
 
Therefore, in a gauge fixed formulation, one needs to add to the 10D action an appropriate counterterm compensating for the non-trivial transformation of the fermion path integral measure under \eqref{GeneralPsiTransf}. The quantum theory is expected to be symmetric only under the discrete subgroup $\sz\subset\sr$,\footnote{Taking into account the action on fermions, the group $\sz$ should be replaced by a non-trivial $\mathbb{Z}_2$ extension thereof \cite{Pantev:2016nze}. However, we will ignore this subtlety here, as our focus is on local gauge anomalies and hence on loops which are not able to detect any global sign ambiguities.} and hence anomaly cancellation requires a Chern-Simons-like counterterm with suitable modular properties\footnote{As explained in \cite{Gaberdiel:1998ui}, anomaly cancellation is not enough to completely fix the modular function of the counterterm. Here we adopt the choice proven in \cite{Gaberdiel:1998ui} to be consistent with compactications of the 8D theory.}, such as
\begin{equation}
  \label{GGterm}
  S^{\rm (10)}\;\supset\;i\int ln\left( \frac{\eta(\tau)\,\bar{j}^{1/12}(\bar{\tau})}{\bar{\eta}(\bar{\tau})\,j^{1/12}(\tau)}\right)\left[2 X_8^-(R)+\frac{p_1(R)}{48}\left(\frac{F}{2\pi}\right)^2-\frac{1}{32}\left(\frac{F}{2\pi}\right)^4\right]\, \frac{F}{2\pi}\,.
  \end{equation} 

In the absence of 7-branes there is no anomaly and the above coupling is completely inert under $\sz$: Indeed $F=dQ$ globally and the integrand of \eqref{VarPathInt} can be easily seen to reduce to a total derivative.

In the presence of 7-branes, instead, $F$ represents a non-trivial cohomology class. The equations of motion for the background link this quantity to the metric of space-time as
\begin{equation}\label{EOM}
F=-\frac{i}{2}trR\,,
 \end{equation}
so that the non-trivial vacuum profile for the axio-dilaton induces a non-trivial Ricci curvature. The $F^5$-term in \eqref{GGterm} would only contribute if we allow $\langle\tau\rangle$ to vary over the whole 10D space. In the language of F-theory such a situation would mean compactifying on an elliptically fibred Calabi-Yau \emph{sixfold}. While there is no reason to believe that this is inconsistent, we would soon run into trouble due to the lack of a fundamental definition of F-theory. The cubic and linear terms of \eqref{GGterm}, instead, are both present for instance in a compactification to 4D, and it is an interesting question to find their F/M-theory lift. This lift should be written solely in terms of gravitational invariants of the elliptic Calabi-Yau fourfold (in the same spirit of \cite{Minasian:2015bxa}). Our present goal, however, is more modest. We restrict our attention to space-times of the form $\mathcal{M}_{10}=\mathcal{M}_{8}\times S^2$, where $\langle\tau\rangle$ is only allowed to vary on the 2-sphere, and study the consequences of the $F$-linear part of the coupling \eqref{GGterm}, which is the only one that survives in this case.

\subsection{F-theory on K3 and 7-brane couplings}

As is well known, F-theory compactifications on K3 are a class of type IIB vacua preserving minimal supersymmetry in 8D and involving exactly 24 vortex-like sources (7-branes) localized on the 2-sphere. It is also well known that not all of these 7-branes are mobile, due to certain global obstructions which leave only at most 18 of them free to move around. It is amusing to realize that the structure of the Green-Gaberdiel anomaly, reviewed in the previous section, ``knows" about these gravitational constraints, in a sense that we are now going to explain.

By a suitable choice of $\sz$ frame, one can choose the 18 dynamical sources to be ordinary D7-branes which in a generic region of the moduli space give rise to a $U(1)^{18}$ gauge group. The resulting 8D theory is an $N=1$ supergravity coupled to 18 Maxwell supermultiplets. The moduli space of this theory includes the coset $\frac{SO(2,18)}{SO(2)\times SO(18)}$, and the couplings of the various fermions to the $SO(2)$ are chiral \cite{Salam:1984ft}. The theory therefore suffers from a local $U(1)$ anomaly \cite{Marcus:1985yy} whose cancellation mechanism is very much analogous to the 10D one previously reviewed. There is however a crucial difference with respect to the Green-Gaberdiel anomaly: The 8D anomaly is not related to sources of monodromy for the moduli fields, but it is there even in their absence. The reason is that the 8D anomaly polynomial is a 10-form including a term linear in the composite field strength (and quartic in the Riemann tensor), which need not vanish.
We will have much more to say about this class of anomalies in section \ref{sec:N=1}, where we will analyze them for a variety of vacuum configurations with non-Abelian gauge groups. We will assume absence of branch cuts for the 8D moduli, with the exception of a few remarks in sections \ref{sec:glo} and \ref{sec:DO}.

We leave the general discussion to section \ref{sec:N=1}, where we list the fermionic fields of these theories, together with their $U(1)$ charges and chiralities. Here we just state the result for the case at hand: Taking into account the contributions of the gravitino, the dilatino and the 18 gaugini, the ensuing anomaly is cancelled by adding the following counterterm to the 8D effective action
\begin{equation}\label{AnoU(1)18}
S^{\rm (8)}\;\supset\; \int f({\bf z},{\bf\bar{z}})\, \frac{1}{32(2\pi)^4}\left[\frac{11}{15}trR^4 - \frac{1}{12}(trR^2)^2\right]\,,
\end{equation}
where $f(\bf{z},\bf{\bar{z}})$ is a function of the moduli, collectively denoted by $\bf{z}$, with the appropriate modular properties to counterbalance the anomalous phase variation in the path integral.
Remarkably, the quartic polynomial in the 8D Riemann tensor appearing in \eqref{AnoU(1)18} is exactly reproduced by adding the contribution of 24 punctures to the polynomial \eqref{VarPathInt} fixed by the 10D anomaly cancelation. More precisely, one first brings the 10D anomalous  phase variation down to 8D, by using the fact that $F/2\pi$ integrates to $-2$ on the 2-sphere, and that all its higher powers vanish. This is due to the relation \eqref{EOM}, which means that $-F/2\pi$ represents the first Chern class of the tangent bundle of the string internal manifold \cite{Greene:1989ya}. Then one adds a term due to the 24 punctures, as if each of them would contribute a dynamical gaugino of charge $1/2$. All in all one obtains:
\begin{eqnarray}
4X_8^-(R)+24\times\frac{1}{2}\widehat{A}(\mathcal{M}_8)|_{\rm 8-form}&=& \frac{1}{32(2\pi)^4}\left[\frac{11}{15}trR^4 - \frac{1}{12}(trR^2)^2\right]\,,
\end{eqnarray}
where $\widehat{A}(\mathcal{M}_8)$ is the so-called A-roof genus, quoted in \eqref{DiracIndex}. \\

We would now like to point out another intriguing implication of the 10D $\sz$ anomaly cancellation. We will indeed argue that the Green-Gaberdiel counterterm \eqref{GGterm} codifies the structure of higher-derivative $R^4$ couplings on D7-brane worldvolumes in the regime of strong string coupling\footnote{Note that $R^2$ couplings on the D7-brane trivially extend to strong $g_s$, as they induce D3-brane charge which is S-duality invariant. See \cite{Grimm:2013gma,Grimm:2013bha} for their F/M-theory origin.}. The story is analogous to the one of $R^2$ couplings on D3-branes \cite{Bachas:1999um}, whose expression for any value of the string coupling is dictated by the cancelation of an $\sz$ anomaly of the $N=4$ Maxwell theory living on the D3-brane. Here, however, things are more involved, as D7-branes are not singlets under $\sz$. But F-theory teaches us how to handle this problem: As long as gravitational effects are concerned, at strong coupling the physics of D7-branes is completely codified by a non-Ricci-flat 10D bulk geometry together with a non-trivial axio-dilaton profile. Therefore from this perspective it is very reasonable to look at 10D $\sz$ anomaly cancelation to seek for the strong coupling completion of $R^4$ terms on D7-branes. In the following we provide strong evidence that the counterterm \eqref{GGterm} plays the role of such a completion.

To this end, we first take a weak string coupling limit $g_s=\langle\tau_2\rangle^{-1}\to0$ of \eqref{GGterm} and bring the coupling down to 8D by using, as before, $\int_{S^2}F/2\pi=-2$ (all higher $F$-powers being zero). We thus obtain:
\begin{equation}\label{8DGG}
2\pi\int\tau_1\,X_8^-(R)\,,
\end{equation}
where we have used that in this limit $\;ln\left( \frac{\eta(\tau)\,\bar{j}^{1/12}(\bar{\tau})}{\bar{\eta}(\bar{\tau})\,j^{1/12}(\tau)}\right)\to\frac{i\pi\tau_1}{2}$.

Let us now compare \eqref{8DGG} with the weak coupling expectation of the higher-derivative couplings to the Ramond-Ramond axion $\tau_1$. To do that we have to compute the total D(-1)-brane charge induced by the brane content of the theory. In a regime of weak coupling the 24 7-branes arrange themselves in 4 O7$^-$-planes and 16 D7-branes plus 16 D7-images \cite{Sen:1996vd} (see also \cite{Esole:2012tf}). Since an integral (mobile) D(-1)-brane charge is made up of a pair D(-1)/image-D(-1) brane, we must compute it on the orientifold double cover of the 2-sphere. We use the well known formulae for the induced brane charges \cite{Green:1996dd, Cheung:1997az, Minasian:1997mm}, which for a single Dp-brane (with trivial normal and gauge bundle) read $\Gamma_{-1}^{\rm Dp}=2\pi \sqrt{\widehat{A}(\mathcal{M}_8)}$ and for a single O7$^-$-plane (with trivial normal bundle) read\footnote{The symbol $\mathcal{M}_8/4$ means taking $1/4$ of $R$, the curvature of the tangent bundle  \cite{Morales:1998ux, Stefanski:1998he}.} $\Gamma_{-1}^{\rm O7}=-16\pi \sqrt{\widehat{L}(\mathcal{M}_8/4)}$ (see appendix \ref{sec:index} for the relevant definitions). In addition to that, there is a density of D3-brane charge which (if part of $\mathcal{M}_8$ is compactified) needs to be added to cancel the one induced by the 24 7-branes. This amounts to $p_1(R)/2$. Of course these D3-branes also induce D(-1)-brane charge and, if we take into account that too, we obtain the following axion coupling:
\begin{eqnarray}\label{D(-1)coup}
& &\int\tau_1\,\left(32\times\Gamma_{-1}^{\rm D7}\,+\,4\times\Gamma_{-1}^{\rm O7}\,+\,\frac{p_1(R)}{2}\times\Gamma_{-1}^{\rm D3}\right)\\\nonumber
&=&\frac{2\pi}{192 (2\pi)^4}\int\tau_1\,\left(32 \times \frac{1}{32 \times 15}\left[ 8 trR^4 + 5 (trR^2)^2\right]-4\times \frac{1}{16 \times 15}\left[5(trR^2)^2 -28 trR^4 \right] -\frac{(trR^2)^2}{2}\right)\\\nonumber
&=&\frac{2\pi}{192 (2\pi)^4}\int\tau_1\,\left(trR^4 + \frac{1}{4}(trR^2)^2 -\frac{(trR^2)^2}{2}\right)\\\nonumber
&=&2\pi\int\tau_1\,X_8^-(R)\,,
\end{eqnarray}
i.e. exactly what is predicted by the 10D $\sz$ anomaly cancelation.

This remarkable match comes with an annoying puzzle which remains to be explained: Why should the F-theory coupling \eqref{GGterm} ``know" about D3-branes, which do not backreact on the axio-dilaton. Notice that the D3-brane contribution, i.e. the last piece in the l.h.s. of \eqref{D(-1)coup}, just flips the sign of $(trR^2)^2$ in \eqref{X8pm} from $+$ to $-$. The sign flip could presumably be explained in an alternative way, by a suitable redefinition of the Ramond-Ramond four-form potential $C_4$. Such a redefinition, at any value of the string coupling, should look like
\begin{eqnarray}
C_4&\longrightarrow& C_4+\frac{i\,\alpha'^2}{192(2\pi)}\,ln\left(\frac{j(\tau)}{\bar{j}(\bar{\tau})}\right)\,\frac{tr R^2}{(2\pi)^2}\,,
\end{eqnarray}
which respect the $\sz$ invariance of $C_4$. Operating this redefinition adds an additional contribution to the induced D(-1)-brane charge on D7-branes and explains the sign flip from $-$ to $+$ when going to weak coupling, without relying on explicitly added D3-branes. We hope to clarify this issue in the future.

\section{The $\sz$ anomaly in D=8, N=2 supergravity}\label{sec:N=2}
We now turn to the case of $N=2$ supergravity in 8D, which is obtained, for instance, by a $T^2$ compactification of the type IIB theory. We will discuss the structure of the anomaly counterterm and match it with the higher-curvature terms inferred from string amplitudes.

Let $\U$ be the complex structure and $\T$ be the (complexified) K\"ahler structure of $T^2$ respectively. The moduli space of the theory is 
\begin{equation}
\frac{SL(2,\mathbb{R})}{U(1)} \times \frac{SL(3,\mathbb{R})}{SO(3)}\,,
\end{equation}
where the first factor is parametrized by $U$.

The field content of this 8D theory is given by a graviton supermultiplet comprising \cite{Salam:1984ft,Basu:2011he}:  1 graviton,  2 gravitini (doublet under $Spin(3)=SU(2)$),  6 vectors, 2+4 dilatini (doublet + quadruplet under $Spin(3)=SU(2)$), 7 real scalars, 3 2-forms and 1 3-form. The U(1) charges of the gravitini, of the doublet of dilatini and of the quadruplet of dilatini are respectively (they are all positive chiral): $\frac{1}{2}$, $\frac{3}{2}$ and $-\frac{1}{2}$. Finally, the 4-form field strength can be split in self-dual and anti-self-dual part, carryiing charges $1$ and $-1$ respectively under $U(1)$ \cite{Basu:2011he,Bossard:2014lra, Marcus:1985yy}. Hence, using the index formulae in appendix \ref{sec:index}, the 10-form anomaly polynomial is given by \cite{Bossard:2014lra}
\begin{equation}  
P_{\rm 10}= \frac{F}{2\pi}\, \left[2\times \frac{1}{2} I^{d=8}_{3/2}-4\times \frac{1}{2} I_{1/2}+ 2\times \frac{3}{2} I_{1/2} + 2\times I_{\rm form} \right]_{\rm 8-form}\,,
\end{equation}
where, in analogy to \eqref{CompFS}, $F$ is the composite field strength built out of $U$. By the descent method we thus deduce the following anomalous phase variation of the path integral
  \begin{equation}
  \Delta= - 12\int \Sigma\; X_8^-(R) \,,
  \end{equation}
where $X_8^-(R)$ is defined in \eqref{X8pm}.  
As in section \ref{sec:green-gaberdiel}, gauge fixing translates the $U(1)$ anomaly into an  $SL(2,\mathbb{Z})$ anomaly, which we can cancel by introducing in the 8D action a counterterm of the form\footnote{Throughout this paper we neglect possible, subtle moduli-independent shifts of the counterterms, like those considered in \cite{Gaberdiel:1998ui}.} 
 \begin{equation}
  \label{d8n2}
  S_{\rm IIB}^{(8)}\supset 12 i\int ln\left( \frac{\eta(U)}{\bar{\eta}(\bar{U})}\right)  X_8^-(R)\,.
 \end{equation}

In the case of type IIA compactified on $T^2$ we get by T-duality ($T\leftrightarrow U$)
\begin{equation}\label{d8n21}
S_{\rm IIA}^{(8)}\supset 12 i\int ln\left( \frac{\eta(T)}{\bar{\eta}(\bar{T})}\right)  X_8^-(R)\,.
\end{equation}
The above terms \eqref{d8n2} and \eqref{d8n21} are consistent with string amplitude results. Indeed, if we extract the CP-odd part of the 8D effective action from the string amplitude computations of \cite{Kiritsis:1997em}, we get exactly the higher-derivative structure of \eqref{d8n2} and \eqref{d8n21}.

As a final remark, note that the terms \eqref{d8n2} and \eqref{d8n21} also have the correct properties in the decompactification limit, in which we send the volume of $T^2$ to infinity. Indeed, while the coupling \eqref{d8n2} vanishes in this limit (see appendix \ref{sec:decomp} for the details), the one in \eqref{d8n21} gives rise to the well-known higher-curvature terms in type IIA/M theory \cite{Vafa:1995fj, Duff:1995wd}.\footnote{In fact, taking a radius of a single circle to infinity, one finds a  nine-dimensional coupling $A_1 X_8^-(R)$, where in IIA frame $A_1 = B_{\mu 9} dx^{\mu}$, and in IIB, $A_1 = \frac{\alpha'}{R_9^2} g_{\mu 9} d x^{\mu}$, where $R$ is the radius of the remaining circle. In IIB, the coupling is subsequently suppressed in the ten-dimensional  limit. In IIA/M-theory it eventually lifts to  $B_2 X_8^-(R)$ / $C_3 X_8^-(R)$, unifying the composite $U(1)$ anomaly in 8D and M5-brane anomaly.}

\section{The $\sz$ anomaly in D=8, N=1 supergravity}\label{sec:N=1}
Let us now turn our attention to minimal supergravity in 8D and its possible anomaly counterterms. We obtain this type of theories by compactifying the Heterotic theories (either $SO(32)$ or $E_8\times E_8$) down to 8D on a $T^2$. We will focus in particular on compactifications where the gauge group, left unbroken, does not contain $U(1)$ factors. For the various cases, we will first compute the $\sz$ counterterm needed to cancel the anomaly; then, by computing the relevant 5-point string amplitudes, we will be able to show that their leading harmonic CP-odd part exactly reproduces the higher-derivative structure of the counterterm.  For compactifications which break the original $SO(32)$ or $E_8\times E_8$ gauge group to $G$, additional higher-derivative structures arise from the amplitude, due to the contribution from the massive vector-multiplets: We will show that such new structures fit nicely with the group theory decomposition under $SO(32)\rightarrow G$ or $\ett\rightarrow G$.

\subsection{Generalities}
\label{sec:d8gen}
Generically, $T^2$ compactifications of the Heterotic theories have in their spectrum a gravity multiplet comprising 1 graviton, 1 antisymmetric 2-form, 1 gravitino, 2 graviphotons, 1 dilatino, 1 real scalar, and $n$ vector multiplets comprising $n$ photons, $n$ gaugini and $2n$ real scalars parametrizing the coset $\son$. For specific values of the Wilson lines, the unbroken gauge group does not contain $U(1)$ factors, and correspondingly we get only $n=2$ abelian vector multiplets. Their complex scalars $T,U$ are identified with K\"ahler and complex structure of $T^2$ respectively, and parametrize the coset $\frac{SO(2,2)}{U(1)\times U(1)}$.

The fermions of the theory have chiral couplings to one of the $U(1)$'s of the coset \cite{Salam:1985ns}. Since in this type of compactifications there is an exchange symmetry\footnote{More precisely, these theories are invariant under the descrete group $O(2,2,\mathbb{Z})=\sz_T\times\sz_U\rtimes\mathbb{Z}_2$ to all orders in string perturbation theory \cite{Antoniadis:1995ct} (see also \cite{GarciaEtxebarria:2012zm}). In this paper we will not care about the T-duality properties of the various couplings we discuss, but will only pay attention at their higher-derivative structure.} between $T$ and $U$, we can discuss the ensuing $\sz$ anomaly in terms of either of these two moduli, and there will be a corresponding counterterm involving the other modulus as well. The $U(1)$ charges of the gravitino (positive chirality), the dilatino (negative chirality) and the gaugini (positive chirality) are all $\frac{1}{2}$ (we are working with Weyl representations).

Using the index formulae in appendix \ref{sec:index}, we find the following anomalous phase variation of the path integral\footnote{All traces $Tr$ with a capital $T$ are in the adjoint representation of the group $G$.}
\begin{equation}
  \Delta_G=-\int\frac{\Sigma}{32(2\pi)^4}\left[(248+{\rm dim}G)\left[\frac{trR^4}{360}+\frac{(trR^2)^2}{288} \right] -(trR^2)^2 + \frac{1}{6}trR^2 TrF^2 + \frac{2}{3}TrF^4\right].
  \end{equation}

 As in all previous cases, gauge fixing induces an $\sz$ anomaly, which is canceled by a coupling in the 8D effective action of the type
 \begin{equation}\label{AnoGen}
 S^{\rm (8)}\;\supset\; \int f({\bf z},{\bf\bar{z}})\,Y_8^G\,,
 \end{equation}
 where ${\bf z}$ is either $T$ or $U$, $f$ is a function with the appropriate modular properties required by anomaly cancelation\footnote{As in \cite{Gaberdiel:1998ui}, anomaly cancellation does not uniquely fix the function $f$. A possible form of this function may be given by $f(z,\bar{z})=log\frac{\eta^{24}(z)}{\bar{\eta}^{24}(\bar{z})}+log\frac{j(z)}{\bar{j}(\bar{z})}$ so that it cancels the $\sz$ variation and does not decompactify to 10D as this anomaly term is local to 8D theory.}, and we have defined
\begin{equation}
  \label{n1pol}
  Y_8^G=\frac{1}{32 (2\pi)^4}\left[ (248+{\rm dim}G)\left[\frac{trR^4}{360}+\frac{(trR^2)^2}{288} \right]-(trR^2)^2 + \frac{1}{6}trR^2 TrF^2 + \frac{2}{3}TrF^4\right].
  \end{equation}
By explicit calculation, we will show that the term \eqref{AnoGen} is reproduced in the CP-odd sector of 5-point one-loop string amplitudes with 1 modulus and either 4 gravitons, or 4 gauge bosons, or 2 gravitons and  2 gauge bosons. Because of this relationship to anomalies, this term does not receive any further renormalization from higher string loops. We remark that these counterterms do not lift to 10D in the decompactification limit, because the massless particles responsible for the anomaly are different in the two theories.

In the following subsections we will explore the anomaly structure for the theories with $G= SO(32), \,\ett, \,SO(16)^2$  and  $SO(8)^4 $.

\subsection{$\bf SO(32)$ and $\bf E_8 \times E_8$}\label{sec:G32}
In the cases of $G=SO(32)$ and $G=\ett$ the 8-form polynomial in \eqref{n1pol} takes the form
   \begin{subequations}
   \begin{align}
   Y_8^{SO(32)}&=\frac{1}{32 (2\pi)^4}\left( \frac{31}{15} trR^4+\frac{19}{12}(trR^2)^2+5trR^2 trF^2+2(trF^2)^2+16trF^4\right),\label{32pol}\\
   Y_8^{\ett}&=\frac{1}{32 (2\pi)^4}\left( \frac{31}{15} trR^4+\frac{19}{12}(trR^2)^2+5trR^2 \sum_{i=1}^2trF_i^2+6\sum_{i=1}^2(trF_i^2)^2\right),\label{ettpol}
   \end{align}   
   \end{subequations}
where we have converted the traces in the adjoint into traces in the fundamental representation (indicated by the lower-case symbol $tr$), using the formulae in \eqref{traces}.   
   
Let us now compare the higher-derivative structures \eqref{32pol}, \eqref{ettpol} with the $SO(32)$ and the $\ett$ 5-point string amplitudes which have been calculated in \cite{Kiritsis:1997hf, Harvey:1995fq}. Here we just give the results. For the details we refer to \cite{Kiritsis:1997hf, Harvey:1995fq, Sasmal:2016fap}.

For the case $G=SO(32)$, the amplitude gives rise to the following $\alpha'^3$ corrections in the 8D effective action\footnote{Note that $\T$.}
  \small
  \begin{eqnarray}
  \label{32action}
  S_{\text{amp}}^{SO(32)}&=&\frac{1}{192(2\pi)^3}\int B_{89}\underbrace{\left(trR^{4}+\frac{1}{4}(trR^2)^{2}+trR^{2} trF^{2} + 8trF^{4} \right)}_{\text{GS}}\nonumber\\
   &{ }& +\frac{1}{4 \times 192(2\pi)^4 }\int\left[\leou \right]\times\nonumber\\
   &{}&\underbrace{\left(\frac{31}{15} trR^{4}+\frac{19}{12}(trR^2)^{2}+5trR^{2} trF^{2}+2(trF^2)^{2}+16trF^{4} \right)}_{\text{anomaly term for } U}\nonumber\\
  &{}& +\frac{1}{4 \times 192(2\pi)^4}\int\left[\leot -4i\pi T_1\right]\nonumber\\
  &{}& \times\underbrace{\left(\frac{31}{15} trR^{4}+\frac{19}{12}(trR^2)^{2}+5trR^{2} trF^{2}+2(trF^2)^{2}+16trF^{4} \right)}_{\text{ anomaly term for } T}.
  \end{eqnarray}
  \normalsize
  In the above, the term in the first line, comes from the trivial orbit part of the $\Gamma_{2,2}$ lattice in the partition function of the Heterotic string compactified on $T^2$ (for the orbit decomposition and related details of string 5-point amplitude, we refer to \cite{Kiritsis:1997hf, Sasmal:2016fap}) and is in fact the $T^2$ reduction of the Green-Schwarz term of the 10D $SO(32)$ Heterotic theory \cite{Green:1984sg}
  \begin{equation}
S_{GS}=\frac{1}{192(2\pi)^5 \alpha '}\int B_2 \wedge \left(trR^{4}+\frac{1}{4}(trR^2)^{2}+trR^{2} trF^{2} + 8trF^{4} \right)\,.
\end{equation}
 
Remarkably, the remaining two terms in \eqref{32action} involve an 8-form polynomial which exactly matches the one in \eqref{32pol} predicted by anomaly cancelation. As expected from T-duality, one is the counterterm for the $U$ modulus and the other is the counterterm for the $T$ modulus. Notice moreover, as already mentioned, that both this terms disappear in the decompactification limit to 10D $V_{T^2}\to0$ (see appendix \ref{sec:decomp} for some details).
This match is due to the fact that the low-energy limit of the 5-point 1-loop string amplitude is the 1-loop amplitude in supergravity, and that the IR divergence in the string loop amplitude manifests itself in the quantum anomaly of the low-energy effective theory, detected from the UV divergence of the supergravity 5-point 1 loop amplitude \cite{Dixon:1990pc}.

Similarly, for the case $G=\ett$ we find 
  \begin{eqnarray}
  \label{e88action}
  S_{\text{amp}}^{\ett}&=&\frac{1}{192(2\pi)^3}\int B_{89}\underbrace{\left( trR^4+\frac{1}{4}(trR^2)^2+trR^2  \sum_{i=1}^2trF_i^2-2trF_1^2trF_2^2+2\sum_{i=1}^2(trF_i^2)^2\right)}_{\text{GS term = Trivial orbit amplitude}} \nonumber\\
   &{ }& +\frac{1}{4 \times 192(2\pi)^4 }\int\left[\leou \right]\times\nonumber\\
   &{}& \underbrace{\left( \frac{31}{15} trR^4+\frac{19}{12}(trR^2)^2+5trR^2 \sum_{i=1}^2trF_i^2+6\sum_{i=1}^2(trF_i^2)^2\right)}_{\text{ anomaly term for } U}\nonumber\\
  &{}& +\frac{1}{4 \times 192(2\pi)^4}\int\left[\leot -4i\pi T_1\right]\nonumber\\
  &{}& \times\underbrace{\left( \frac{31}{15} trR^4+\frac{19}{12}(trR^2)^2+5trR^2 \sum_{i=1}^2trF_i^2+6\sum_{i=1}^2(trF_i^2)^2\right)}_{\text{ anomaly term for } T}.
  \end{eqnarray}
 Also in this case the trivial orbit part matches the $T^2$ reduction of the 10D $\ett$ Green-Schwarz term \cite{Green:1984sg}
  \begin{equation}
   \label{e88gs}
S_{GS}=\frac{1}{192(2\pi)^5 \alpha '}\int B_2  \left( trR^4+\frac{1}{4}(trR^2)^2+trR^2 \sum_{i=1}^2trF_i^2-2trF_1^2trF_2^2+2\sum_{i=1}^2(trF_i^2)^2 \right),
\end{equation}
whereas the other two pieces reproduce the anomaly polynomial \eqref{ettpol} expected from anomaly cancelation.


\subsection{$\bf SO(16)^2$}\label{sec:G16}
We now consider 10D $\ett$ Heterotic string theory compactified on $T^2$ with the following Wilson line configuration
\begin{equation}
Y_i^1 = (0^4, \,\frac{1}{2}^4, \,0^4,\, \frac{1}{2}^4), \quad \quad Y_i^2= (0^8,\,0^8),\qquad i=1,\cdots ,16,
\end{equation}
so that the gauge group is broken to $SO(16) \times SO(16)$ in 8D. One can of course rearrange the 8 non-zero values of the Wilson lines so that one can start from the $SO(32)$ gauge group in 10D and again obtain $SO(16) \times SO(16)$ in 8D.

Adapting formula \eqref{n1pol} to the adjoint representation $\textbf{(120,1) $\oplus$ (1,120)}$ of $SO(16)^2$ and using the trace-conversion formulae \eqref{traces} we can write the 8-form polynomial for the case $G=SO(16)^2$ as follows
\begin{equation}\label{so162pol}
Y_8^{SO(16)^2}=\frac{1}{32 (2\pi)^4}\left( \frac{488}{360} trR^4+\frac{200}{288}(trR^2)^2+\frac{7}{3}trR^2  \sum_{i=1}^2trF_i^2 +\frac{16}{3} \sum_{i=1}^2trF_i^4+2 \sum_{i=1}^2(trF_i^2)^2\right).
\end{equation}

We will now compare this supergravity result with the string amplitude \cite{Gutperle:1999dx, Sasmal:2016fap}. To this end it is convenient to decompose the string partition function in three $\sz$ orbits: trivial, degenerate and non-degenerate (see \cite{Sasmal:2016fap} for the details). The final result is recorded here for reader's convenience
\begin{align}
  \label{1610}
  \mathcal{A}_{\text{trivial}}&= T_1\left( trR^4+\frac{1}{4}(trR^2)^2+trR^2  \sum_{i=1}^2trF_i^2-2trF_1^2trF_2^2+2 \sum_{i=1}^2(trF_i^2)^2\right)\,,
  \end{align}
  
\begin{align}
  \label{1620}
  \mathcal{A}_{\text{deg.}}=\frac{1}{96(2\pi)^4}\leou &\left( \frac{488}{360} trR^4+\frac{200}{288}(trR^2)^2+\frac{7}{3}trR^2  \sum_{i=1}^2trF_i^2 \right.\nonumber\\
  &{}\left.+\frac{16}{3} \sum_{i=1}^2trF_i^4+2 \sum_{i=1}^2(trF_i^2)^2\right)\,,
  \end{align}  
 
 \begin{subequations}
 \begin{align}\label{1630a}
 \mathcal{A}_{\text{non-deg.}}=&\frac{1}{96(2\pi)^4} \left[ 8i\pi T_1 -ln\left( \frac{\eta^{24}(2T)}{\bar{\eta}^{24}(2\bar{T})}\right) \right] \left( \frac{488}{360} trR^4+\frac{200}{288}(trR^2)^2 +\frac{7}{3}trR^2  \sum_{i=1}^2trF_i^2\right.\nonumber\\
 &\hspace{5.5cm}\left. +\frac{16}{3} \sum_{i=1}^2trF_i^4+2 \sum_{i=1}^2(trF_i^2)^2\right)+\\ \nn \\
&+ \frac{1}{192(2\pi)^4}\left[-4i\pi T_1 +  ln\left( \frac{\eta^{24}(2T)}{\bar{\eta}^{24}(2\bar{T})}\right)-ln\left( \frac{\eta^{24}(T)}{\bar{\eta}^{24}(\bar{T})}\right) \right]\times\nn\\
&\times\left[256\left( \frac{trR^4}{360} + \frac{(trR^2)^2}{288}\right) +\frac{8}{3}trR^2 \sum_{i=1}^2trF_i^2-\frac{16}{3}\sum_{i=1}^2trF_i^4 +4\sum_{i=1}^2(trF_i^2)^2\right]\label{1630b}\,.
 \end{align}
 \end{subequations}
 
The above expressions translate into the following $\alpha'^3$ corrections to the 8D effective action
 \begin{equation}
  \label{164}
  S_{\text{amp}}^{SO(16)}=\int \left[ N_1 \mathcal{A}_{\text{trivial}}+ N_2 \mathcal{A}_{\text{deg.}}+N_3\mathcal{A}_{\text{non-deg.}}\right]\,,
  \end{equation}
with appropriate normalization factors $N_1, ~N_2, ~N_3$. As usual, the trivial orbit term \eqref{1610} is the same as the $T^2$ reduction of the Heterotic Green-Schwarz term, as can be most easily seen by starting from the $\ett$ one in \eqref{e88gs}. The degenerate orbit term \eqref{1620} and the first piece of the non-degenerate orbit term \eqref{1630a} involve exactly the same 8-form polynomial dictated by anomaly cancelation \eqref{so162pol}.\footnote{Note that the modular coefficients in front do not appear to have the correct modular properties to cancel the corresponding anomalous $\sz$ phase variations, but this is only an artefact of having split the amplitude in three different orbits and of having adopted appropriate renormalization schemes \cite{Kiritsis:1997hf}.} Finally, the second piece of the non-degenerate orbit term \eqref{1630b} is the contribution from the massive vector multiplets in the $\textbf{ (128,1) $\oplus$ (1,128)}$ representation of $SO(16)^2$, as can be verified using the trace formulae \eqref{16trace1}, \eqref{16trace2}.

One can of course make a totally analogous analysis starting from $SO(32)$, and using the branching rules \eqref{so32BR} and the trace formulae \eqref{16trace}.
 
\subsection{$\bf SO(8)^4$}\label{sec:G8}
The last case we analyze is the 8D N=1 theory with $G=SO(8)^4$,  which can again be obtained from either Heterotic string theory compactified on $T^2$ with appropriate Wilson lines along the two 1-cycles of the torus. Adapting formula \eqref{n1pol} to the adjoint representation of $SO(8)^4$ and using the trace-conversion formulae \eqref{traces}, the 8-form polynomial for the case $G=SO(8)^8$ gets the relatively simple form
\begin{equation}
\label{so8u1}
Y_8^{SO(8)^4}=\frac{1}{32(2\pi)^4} \left( trR^4+\frac{1}{4}(trR^2)^2+trR^2 \sum_{i=1}^4 trF_i^2+2\sum_{i=1}^4 (trF_i^2)^2\right)\,.
\end{equation}

Before moving to the string amplitude, we note that the expression \eqref{so8u1} has a very interesting rewriting
\begin{equation}
\label{HW}
Y_8^{SO(8)^4}= \frac{1}{32(2\pi)^4}\left[\left( trR^4-\frac{1}{4}(trR^2)^2\right) +\frac{1}{2}\sum_{i=1}^4 \left( \frac{1}{2}trR^2+ 2trF_i^2\right)^2 \right] \,.
\end{equation}
\normalsize
In \eqref{HW} we recognize the first piece as (proportional to) the M5-brane anomaly polynomial $X_8^-(R)$, that we already met in the type IIB context in section \ref{sec:green-gaberdiel} (see \eqref{X8pm}). The second piece of \eqref{HW}, on the other hand, being a sum of squares, strongly suggests that an Ho\v{r}ava-Witten-like mechanism is at work here \cite{Horava:1996ma}. We hope to come back to this intriguing observation in the future.

Let us now compare our supergravity result with the string amplitude computation \cite{Gutperle:1999xu, Lerche:1999de, Sasmal:2016fap}. To this end we again decompose the string partition function in three $\sz$ orbits: trivial, degenerate and non-degenerate (details can be found in  \cite{Gutperle:1999xu, Sasmal:2016fap}) with the final result:
\bseq
\small
   \begin{eqnarray}
  \label{81}
&&  \mathcal{A}_{\text{trivial}}= T_1\left( trR^4+\frac{1}{4}(trR^2)^2+trR^2 \sum_{i=1}^4 trF_i^2\right.\\\nn
&&\left.-2trF_1^2trF_3^2-2trF_1^2trF_4^2-2trF_2^2trF_4^2-2trF_2^2trF_3^2 +4trF_1^2trF_2^2+4trF_3^2trF_4^2 +2\sum_{i=1}^4(trF_i^2)^2\right),
  \end{eqnarray}
  \normalsize
  \small
  \begin{eqnarray}
  \label{82}
 & & \mathcal{A}_{\text{degenerate}}+\mathcal{A}_{\text{non-degenerate}}=\nn\\ \nonumber\\
& &\underbrace{\frac{1}{96 (2\pi)^4} \left[ 8i\pi T_1 -ln \left( \frac{\eta(2T)}{\bar{\eta}(2\bar{T})}\right) +\leou\right]\left[ trR^4+\frac{1}{4}(trR^2)^2+trR^2 \sum_{i=1}^4 trF_i^2+2\sum_{i=1}^4 (trF_i^2)^2\right]}_{T\; {\rm and} \;U\; \text{anomaly terms}}\\
\label{83}
&  & + \frac{1}{96 (2\pi)^4} \left[ 8i\pi T_1 -ln \left( \frac{\eta(2T)}{\bar{\eta}(2\bar{T})}\right) +\leou\right]\times\\\nonumber
\label{84}
& & \underbrace{ \left[ 128 \left( \frac{trR^4}{360} + \frac{(trR^2)^2}{288}\right)+\frac{4}{3}trR^2 \sum_{i=1}^4 trF_i^2 + \frac{4}{3} \left(4\sum_{i=1}^4 trF_i^4 +3trF_1^2 trF_2^2 +3 trF_3^2 trF_4^2  \right)  \right]}_{(8,8,1,1)\oplus (1,1,8,8)}\\
& & + \frac{1}{192 (2\pi)^4} \left[ 8i\pi T_1 -ln \left( \frac{\eta(4T)}{\bar{\eta}(4\bar{T})}\right)+\left( \frac{\eta(2T)}{\bar{\eta}(2\bar{T})}\right) +\leou\right]\times\\\nonumber
\label{85}
& & \underbrace{ \left[ 128 \left( \frac{trR^4}{360} + \frac{(trR^2)^2}{288}\right)+\frac{4}{3}trR^2 \sum_{i=1}^4 trF_i^2 + \frac{4}{3} \left(-4\sum_{i=1}^4 trF_i^4 +3\sum_{i=1}^4 (trF_i^2)^2 +6trF_1^2 trF_2^2 +6 trF_3^2 trF_4^2  \right)  \right]}_{(8,8,1,1)'\oplus (1,1,8,8)'}\\
& & + \frac{1}{192 (2\pi)^4} \left[ 8i\pi T_1 -ln \left( \frac{\eta(4T)}{\bar{\eta}(4\bar{T})}\right)+\left( \frac{\eta(2T)}{\bar{\eta}(2\bar{T})}\right) +\leou\right]\times\\\nonumber
\label{86}
& & \underbrace{ \left[ 128 \left( \frac{trR^4}{360} + \frac{(trR^2)^2}{288}\right)+\frac{4}{3}trR^2 \sum_{i=1}^4 trF_i^2 + \frac{4}{3} \left(-4\sum_{i=1}^4 trF_i^4 +3\sum_{i=1}^4 (trF_i^2)^2 +6trF_1^2 trF_2^2 +6 trF_3^2 trF_4^2  \right)  \right]}_{(8,8,1,1)''\oplus (1,1,8,8)''}\\
& &+\frac{1}{24 (2\pi)^4} \left[ 8i\pi T_1 -ln \left( \frac{\eta(2T)}{\bar{\eta}(2\bar{T})}\right) +\leou\right]\times\underbrace{trF_1^2 trF_2^2+trF_3^2 trF_4^2}_{\text{orbifold shifts}}\\
& & +\frac{1}{24 (2\pi)^4}\left[ 8i\pi T_1 -ln \left( \frac{\eta(4T)}{\bar{\eta}(4\bar{T})}\right)+\left( \frac{\eta(2T)}{\bar{\eta}(2\bar{T})}\right) +\leou\right]\times\underbrace{trF_1^2 trF_3^2+trF_2^2 trF_4^2}_{\text{orbifold shifts}} \label{87}\\
& & +\frac{1}{24 (2\pi)^4}\left[ ln \left( \frac{\eta(4T)}{\bar{\eta}(4\bar{T})}\right)-2\left( \frac{\eta(2T)}{\bar{\eta}(2\bar{T})}\right) +\leou\right]\times\underbrace{trF_1^2 trF_4^2+trF_2^2 trF_3^2}_{\text{orbifold shifts}}.\label{88}
   \end{eqnarray}
   \normalsize
   \eseq
The trivial orbit term \eqref{81} is the same as the $T^2$ reduction of the Heterotic Green-Schwarz term, as can be most easily seen by starting from the $\ett$ one in \eqref{e88gs}. The higher-derivative structure \eqref{so8u1} dictated by anomaly cancelation shows up in the term \eqref{82}. The next three terms \eqref{83}, \eqref{84}, \eqref{85} are respectively the contributions from massive vector-multiplets in the representations $(8,8,1,1)\oplus (1,1,8,8)$, $(8,8,1,1)'\oplus (1,1,8,8)'$ and $(8,8,1,1)''\oplus (1,1,8,8)''$ of $SO(8)^4$ (see appendix \ref{Traces&BR} for the notation). This can again be verified most easily starting from $\ett$, and using the branching rule \eqref{e8decomp} and the trace formulae \eqref{8trace}.  Finally, the last three terms \eqref{86}, \eqref{87}, \eqref{88} are due to orbifold shifts, which are generated in the elliptic genus due to the particular combination of Wilson lines.

\section{Global constraints}
\label{sec:glo}

In the remainder of this paper, we will mostly be interested in studying compactifications of the previously discussed 8D theories  on a complex (K\"ahler in fact) manifold $X$, generally with non-vanishing first Chern class. Regardless of supersymmetry, these theories have a composite $U(1)$ connection. As for 10D type IIB, the field strength of the latter will be related to the curvature of the spacetime via an equation analogous to \eqref{EOM}. This equation, together with the fact that the spacetime fermions are charged under the $U(1)$ in question, leads to global constraints and interesting possibilities for possible compactification manifolds.

For the N=2 theory, ironically the general case is simpler: When the duality group is the full $SL(2, \mathbb{R}) \times SL(3, \mathbb{R})$ there is an unambiguous choice of the composite $U(1)$ connection and the whole story closely follows that of 10D IIB theory.\footnote{One may of course think abut the geometrisation of the $SL(3,\mathbb{R})/SO(3)$ and the ensuing U-folds, but this is outside of the scope of our discussion.} However, one may choose to focus on the T-duality subgroup only,\footnote{We choose the perturbative $SL(2,\mathbb{R})$ subgroup of $SL(3,\mathbb{R})$ just for concreteness, but the same discussion should apply to other choices of $SL(2,\mathbb{R})$ subgroups.} i.e. on $SL(2, \mathbb{R})\times SL(2, \mathbb{R})$. Following section \ref{sec:N=2}, we will be thinking of the 8D theory as arising form torus reduction of type IIB  theory, and take  $U$  to be the complex structure modulus and $T$  the complexified K\"ahler modulus. At this point, one may use this accidental second composite $U(1)$, and formulate the global constraints in terms of their sum.\footnote{It is instructive to look at eight-dimensional supersymmetry transformations of the fermions. For example, the gravitino variation is given by 
\bea
\label{eq:grav}
\delta \psi_{\mu} = \left[\nabla_{\mu} -  \frac{i}{4} \frac{\partial_\mu U_1}{U_2} \gamma^9 + \frac14 Q_{\mu}^{ab} T^{ab}\right]\epsilon \,,
\eea
where $T^{ab}$ are $SO(3)$ generators, and the explicit form of the composite connection $Q_{\mu}^{ab}$ as obtained by a reduction from type IIB theory, can be found in \cite{Liu:1997mb}. It is not hard to see that breaking $SO(3)$ to $U(1)$ will make the fermionic derivatives symmetric in the two composite connections. There is however an important difference: Only one of the two $U(1)$ connections comes with $\gamma^9$, i.e. a chiral (anomalous) coupling. The other fermionic variations display similar features.}  Notably the tadpole condition analogous to the 7-brane one in type IIB theory  becomes 
\bea
\label{eq:tadN2}
c_1(TX) + [\Ftau]  + [\Frho]=0\,,
\eea
which is just the direct generalization of the equation \eqref{EOM} written in cohomology.
Note that when looking at possible compactifications on manifolds with nontrivial $c_1$, requiring that either $\Ftau$ or $\Frho$ (or both) are non-trivial in cohomology is a matter of {\sl choice}.  Different choices are a priori consistent and correspond to different backgrounds. We will return to these in section  \ref{sec:DO}.

Let us now turn to N=1 theories. Let the 8D gauge group be $U(1)^n \times G$, where  $G$ is a product of semisimple groups (with $\mbox{rank(}G) = 18-n$). The coset in question is now $SO(2,n, \mathbb{R})/SO(n) \times U(1)$, and we denote the curvature of the composite connection by $F^Q = F(z_i, \bar{z}_i)$, with $z_i$  ($i=1,...,n$) being the complex moduli of the $n$ abelian vector multiplets. The general expression for $F^Q$ is complicated and, contrary to the $N=2$ case, it does not split into a sum of individual terms of the schematic form ${dz_i \wedge d\bar{z}_i}/{(\Im(z_i))^2}$. This remains true even in the absence of Wilson lines (i.e. $n=2$), and is related to the fact that in $N=1$ the coset does not split as opposed to the higher-supersymmetric case.

We will still use $U$ and $T$ for complex stricture and complexified K\"ahler moduli of $T^2$ respectively.  The main difference from N=2 comes for the fact that there are two different tadpole conditions relating $\Ftau$ and $\Frho$ to $c_1(TX)$.  These conditions  can be derived just by thinking about 10D Heterotic strings on elliptically fibered Calabi-Yau manifolds. The first comes from restricting the Calabi-Yau condition of triviality of the canonical bundle to the base via the adjunction formula. The second can most easily be seen as a restriction of the Heterotic Bianchi Identity to the base. Denoting the 10D spacetime (an elliptically fibered space) as $M$, the base of the fibration as $X$ and the gauge bundle as $E$, the Bianchi Identity can be written as
\beq
\label{eq:bi10}
\frac{1}{2}p_1(TM) - c_2(E) - \eta(\mbox{NS5}) = 0\,,
\eeq
where $\eta(\mbox{NS5})$ is the class of the full NS5-brane content. The two conditions on  $X$ are then
\bseq
\bea
\label{eq:bi8}
&& 12 c_1(TX) + c_1({\tilde E})+ 12 [\Frho] = 0\,, \\
\label{eq:bi8*}
&& c_1(TX) + [\Ftau] = 0\,,
\eea
\eseq
where $\tilde E$ is the Abelian part of the 8D gauge group, and, analogously to 7-branes in type IIB, we have defined $[\Frho]=\eta(\mbox{NS5})/12$.  As we will see shortly, the interplay of \eqref{eq:bi8} and \eqref{eq:bi8*} imposes constraints on the choices of allowed backgrounds. For non-geometric backgrounds one may replace $[\Ftau]$ in \eqref{eq:bi8*} by another two-form. However this condition is generally different from \eqref{eq:bi8}, and omitting it leads to anomalous (1,0) theories when compactified on $\mathbb{P}^1$.

This situation appears to be  somewhat different form the previous cases where one could relate $c_1(TX)$ directly to the full composite field strength $F^Q$. But this is indeed true also here.
We recall that K\"ahler manifolds with non-vanishing $c_1$ may be non-spin. As complex manifolds these however always admit a $Spin^c$ structure. As can be seen e.g. from \eqref{eq:grav}, the spinors are ``charged" with respect to the composite $U(1)$, that makes them well defined.\footnote{Note that in the 8D N=2 case, restricting the $SO(3)$ composite connection to $U(1)$ also produces well-defined spinors. In this sense the roles of $T$ and $U$ are symmetric.} In our normalisations this translates into the statement that the phase picked by the spinor upon parallel transport along a closed path is integer, i.e. 
\bea
\label{eq:tadN1}
\frac12 \left(c_1(TX) + [F^Q] \right) = \alpha \, \in H^2(X, \mathbb{Z}) \, .
\eea
In theories with 32 supercharges in 10D and in 8D the two classes are just equal and $\alpha=0$, yielding respectively \eqref{EOM} and \eqref{eq:tadN2}. Here, however, since we are mostly concerned with the two-moduli case in the 8D N=1 theory, equations \eqref{eq:bi8*} and \eqref{eq:bi8} should suffice for the purposes of our next section.

\section{Discussion and outlook}
\label{sec:DO}
The interest in studying eight-dimensional theories is due to the fact that they allow construction and better understanding of more general lower-dimensional string (generally speaking, non-geometrical) backgrounds.
For example,  the torsional heterotic backgrounds realized as principal torus fibrations over $K3$ do not have good 10D large volume limits,  whereas such limit exist in eight dimensions. ``Geometric constructions of non-geometric strings" \`a la \cite{Hellerman:2002ax}   are another example of  how advantageous the 8D descriptions can be.

\vspace{0.3cm}
\noindent
{\bf  N=2 theory.} Here two equally nice ways of putting the theory on a manifold with $c_1(TX) \neq 0$ correspond to taking respectively $\Frho=0$ or $\Ftau=0$ in  \eqref{eq:tadN2}.  Consider the simplest choice of a K\"ahler manifold, $X= \mathbb{P}^1$, yielding six-dimensional backgrounds. The tadpole condition \eqref{eq:tadN2} becomes $\int_X \Ftau = - 2$ or $\int_X\Frho = -2$ for these respective choices.
From the point of view of the IIB theory, the first choice appears to be ``geometric", and it is in fact just a $K3$ compactification of the type IIB strings, while the second is not. The second choice is however geometric for IIA, and corresponds to $K3$ compactification of type IIA strings; conversely the non-trivial $\Ftau$ is not geometrically realized in type IIA theory. Indeed, one would not be getting 6D $(0,2)$ and $(1,1)$ theories from IIA and IIB respectively by means of (geometric) compactification on any manifold. 
An immediate way of seeing why, in spite of the symmetry of \eqref{eq:tadN2}, the two choices result in so markedly different theories in six dimensions is to track the $SL(2,\mathbb{R})_U$ and $SL(2,\mathbb{R})_T$ charges of different fields in the N=2 supergravity. The symmetry between the two is very much broken here.\footnote{The quantum properties of the two $SL(2,\mathbb{R})$ are also very different, resulting in different quantum properties of the 6D $(1,1)$ and  $(0,2)$ theories. The former requires Green-Schwarz-like terms to cancel via inflow chiral string anomalies. The latter is instead completely anomaly free, and has no anomalous higher-derivative couplings.} The bosonic field content is given by
$$
g, \,\, B_2, \,\, \phi, \,\,  (A_2^{T}, \,\,  A_0^{T}), \,\, (( A^+_3)^{U}, \,\, A_1^{U}), \,\, 
A_1^{U, T}, \,\,  U, \,\, T \, , 
$$
where the $U$ and $T$ superscripts indicate with respect to which $SL(2,\mathbb{R})$ the given field transforms (as a doublet). When $\Frho=0$ or $\Ftau=0$, the $p$-form $A_p^T$ or $A_p^U$ respectively is treated as a pair of neutral $p$-forms. When $\Ftau=0$, the doublet of self-dual three-forms simply becomes a single unconstrained three-form field.

In fact both of these choices can be seen as a 8D theory with 24 five-branes (just as F-theory can be thought of as IIB with seven-branes). For non-trivial $T$ modulus these five-branes are the branes of type IIB, i.e. they carry vector multiplets, while for non-trivial $U$ these are the IIA five-branes carrying $(2,0)$ tensor multiplets. 

Let us first look at the $\Ftau=0$ case. In the F-theory context, the mechanism explaining how 24 branes yield only twenty lower dimensional vectors transforming under $SO(2,18)$ is explained in \cite{Douglas:2014ywa}. Here the r\^ole of NS and RR two-forms is played by $A_2^T$. Four more 6D fields come from 
 two {\sl neutral} vector fields of the 8D N=2 theory and $A_3$ (a vector and a three-form which has as many degrees of freedom). These four fields transform under $O(2,2)$ and complete the six-dimensional vectors, giving rise to an $(1,1)$ theory with 20 vector multiplets and symmetry group $SO(4,20)$. One can verify that each of the 20 vectors in the vector multiplets is accompanied by a quartet of scalars as required by supersymmetry. 
 
 For the $\Frho=0$ case, the key field is $(A^+_3)^U$. The mechanism of \cite{Douglas:2014ywa} applies with a shift by one in the rank of the fields. Instead of vectors, one finds  $SO(2,18)$ tensors. To complete the picture for the non-trivial $U$ modulus, we recall that in eight dimensions there are three neutral two-form tensor fields $A_2^T, B_2$, which give rise to three pairs of self-dual and anti-self-dual tensors in six. The resulting $(0,2)$ theory has indeed 21 tensor multiplets and a symmetry group $SO(5,21)$. Once more, one may verify that as required by supersymmetry, each of the 21 anti-self-dual tensors comes with five scalars.

Further examples are beyond the scope of our paper. However, it would be interesting to develop a similar picture for both non-trivial moduli on $X= \mathbb{P}^1$  with $\int_X \Ftau +  \Frho = - 2$, or for higher-dimensional manifolds $X$.

\vspace{0.3cm}
\noindent
{\bf  N=1 theory.} The one-sentence summary of the results of our paper is that the eight-dimensional string amplitudes can be represented as conterterms to the composite $U(1)$ anomaly plus contributions from massive states. We will now point out the crucial importance of this massive sector for the consistency of the lower-dimensional backgrounds. In fact, we will argue that the failure to account for these states properly, i.e. as instructed by string theory, lands the generic 8D N=1 theories in the swampland.

Let us take for the moment the supergravity point of view. Given the eight-dimensional gauge group in the form $U(1)^n \times G$, where $G$ is a product of semisimple groups (with $\mbox{rank(}G) = 18-n$), one would be just coupling this matter to the supergravity in the adjoint. The supergravity will be anomalous under the composite $U(1)$  but a local counterterm can be devised to cancel this anomaly. As we have seen in section \ref{sec:N=1}, for any choice of $G$ other than $E_8 \times E_8$ or $SO(32)$ this conterterm will be different with respect to the effective coupling computed from string amplitudes. In fact, by adding some massive states and integrating these out, one can generate different corrections to the counterterms. The question is if these different (a priori infinitely many) choices are allowed. Ostensibly, all these eight-dimensional theories with proper counterterms are consistent. If however one demands that the theory is consistent on {\sl any} admissible 8D background, the vast majority of these will be ruled out. By taking the eight-dimensional spacetime to be $\mathcal{M}_6 \times \mathbb{P}^1$,  one can show in a large number of cases that all massive completions of D=8 N=1 supergravity with $U(1)^n \times G$ matter except for those obtained from string theory turn out to give rise to anomalous $(0,1)$ theories in six dimensions. 

All our eight-dimensional examples have matter in $U(1)^2 \times G$,  with $\mbox{rank(}G) = 16$, so these will be our main focus here. However we can make some general statements on constraints imposed by \eqref{eq:bi8} and \eqref{eq:bi8*} on string backgrounds.
 
For $n=2$, the Abelian factor $\tilde E$ is missing. Hence, only compactification with $c_1(TX)=0$ or non-geometric compactifications are possible.\footnote{We will call a background geometric, if it can be realized as a compactification on an internal manifold $X$ with non-trivial instanton configurations, but without any extra objects. In particular, Heterotic compactifications with NS5-branes are labeled here as non-geometric.} Equation \eqref{eq:bi8*} allows to build a $K3$ space over $X= \mathbb{P}^1$, whereas equation \eqref{eq:bi8} tells that only NS5-branes participate in cancelling the curvature contribution in the Bianchi Identity. In fact there are 24 of them.

In order to have a geometric Heterotic realisation of D=8 N=1 theory on $X= \mathbb{P}^1$, i.e. a $K3$ compactification,  $n\geq 3$ is required. Indeed taking $[\Frho]=0$, one needs nontrivial $\tilde E$ in order to satisfy \eqref{eq:bi8}. The geometric compactification then should correspond to only $[\Ftau] \neq 0$. Other choices are possible, and these will again lead to non-geometric compactifications.
We hope to return to other cases in a forthcoming publication.\footnote{Different aspects of non-geometric backgrounds using this set-up for low $n$ have been discussed recently in  \cite{McOrist:2010jw, Malmendier:2014uka, Gu:2014ova, Lust:2015yia}.}

Returning to the $n=2$ case, we can check that for $E_8 \times E_8$, the choice $\int_X \Frho = - 2$ leads to an anomaly free $(0,1)$ theory in six-dimensions. Indeed, each five-brane carries a tensor and a hypermultiplet \cite{Duff:1996rs},  and the six-dimensional anomaly factorization condition
\bea
\label{eq:fac6}
244 = N(\mbox{NS5}) (29+1) - \mbox{dim}(G) + 20
\eea
is satisfied for $\mbox{dim}(G) = 496$ and $N(\mbox{NS5}) =24$. The 20 on the r.h.s. is the contribution from 20 neutral hypermultiplets. Note that we have chosen to write the standard factorization condition in terms of the number of five-branes $N(\mbox{NS5})$ rather than of tensor multiplets  (there are 25 of these).
The complete anomaly polynomial is in fact given by
\bea
\label{eq:ano6}
I\propto 2(trR^2)^2+5trR^2 (trF_1^2+trF_2^2)+6((F_1^2)^2+(F_2^2)^2)\,,
\eea
which can be easily rewritten as a sum of 25 factorized terms (as required by \cite{Sagnotti:1992qw}). For $SO(32)$, each five-brane brings along an $USp(2)$ factor with a hyper in bi-fundamental  \cite{Aspinwall:1996nk}, resulting in a factorization condition
\bea
\label{eq:fac6*}
244 = N(\mbox{NS5}) (32-2) - \mbox{dim}(G) + 20 \, .
\eea
It can also be checked that $tr F_{SO(32)}^4$ cancels out and the anomaly polynomial factorises.

For $G=SO(16)^2$ and $G=SO(8)^4$ again $N(\mbox{NS5}) =24$ is required, and there are 20 neutral hypermultiples arising form $K3$, so the formula \eqref{eq:fac6} should formally work. It does however {\sl only}
if vectors in bi-fundamental representations exactly in the form as they appear in sec. \ref{sec:N=1} are included. Hence we see that the correct, i.e. string theoretic, massive completion in eight dimensions, is necessary in order to obtain an anomaly free six-dimensional compactification. In fact we have checked that this applies also for theories with $n \geq 3$, which are outside the scope of this paper.\footnote{Here one also gets hypermultiples charged under the six-dimensional gauge group, equally crucial for the anomaly cancellation mechanism to work.}  Massive states that appear in $T^2$ compactifications to eight-dimensions, must become massless when the torus degenerates to yield an elliptically fibered $K3$. It is worthwhile to understand how this works in detail. It appears that  $G=SO(16)^2$ and $G=SO(8)^4$ should admit a double realization, either with 25 tensor multiplets or with 24 $USp(2)$ gauge fields and hypers in bi-fundamentals. Of course the two versions match once these theories are put on a circle.  To the best of our knowledge the six-dimensional theories with
$G=SO(16)^2$ or $G=SO(8)^4$ gauge groups have not been much discussed in the literature.

Let us end by a comment about the six-dimensional Green-Schwarz term. One can see it arising form integrating the ten-dimensional one on $K3$ (using the constraints imposed by the Bianchi Identity) as in \cite{Duff:1994vv}. The contribution comes only from terms that nicely factorise into internal and external parts as 
\bea
\label{eq:GS8}
B \wedge X_8^{\mbox{\tiny GS}} \rightarrow B \wedge \left[ (a tr R_0^2 + b tr F_0^2) \wedge tr R^2 + (c tr R_0^2 + d  tr F_0^2) \wedge tr F^2 \right]\,.
\eea
Here $R_0$ and $F_0$ are the internal Riemann and gauge field strengths, and $a,b,c,d$ are numerical coefficients.
This form is suggestive of non-vanishing four-point amplitudes in {\sl eight} dimensions involving e.g. the $B$-field, a {\sl composite} $U(1)$ factor and two gravitons. Once more the correct massive sector is important in getting the low-energy contribution matching the six-dimensional Green-Schwarz coupling.

Eight-dimensional N=1 theories hold keys to large classes of interesting string backgrounds, most of which cannot be seen as ordinary compactifications of ten-dimensional string theories. We have argued here, that the study of composite connections and their anomalies may provide useful insights and constraints in constructing these backgrounds.

\section*{Acknowledgments}

We would like to thank  C. Bachas, G. Bossard, I. Garc\'ia-Etxebarria, M. Green, D. Morrison, B. Pioline, A. Sen, S. Theisen, A. Uranga and P. Vanhove for helpful discussions and valuable insights.  RM would like to thank the Korea Institute for Advanced Study and the Simons Center for Geometry and Physics for hospitality during the course of the work. This work was supported in part by the Agence Nationale de la Recherche under the grant 12-BS05-003-01 and CEFIPRA (RM), by the ERC Starting Grant 259133 - ObservableString (RS while at IPhT) and by the ERC Advanced Grant SPLE under contract ERC-2012- ADG-20120216-320421 (RS now).

\newpage
\appendix


\section{Index polynomials}\label{sec:index}

In this Appendix, we list the anomaly polynomials used throughout this article, along with some Chern-Simons forms and their gauge variation forms which come into play during the process of descent \cite{AlvarezGaume:1983ig, AlvarezGaume:1984dr}. All the polynomials are defined on a manifold $\mathcal{M}_d$ of $d=2r$ real dimensions and the rank of the anomaly polynomial is $d+2$, so that we get the anomalous phase variation as an integral over a $2r$-form. But first we give the formalism for the gauge theory: This is important because the reference we are following \cite{AlvarezGaume:1984dr} uses the anti-hermitian generators for the gauge group, while to compute the $U(1)$ anomalies it is useful to work with hermitian generators, as this makes the charges real.
\small
\\
\begin{center}
Table 1: Gauge theory dictionary\\
\(\begin{array}{|c|c|c|c|}
\hline
 \text{Quantities of gauge group G} & \text{Anti-hermitian convention \cite{AlvarezGaume:1984dr}} & \text{Hermitian convention} & \text{Relation} \\
\hline
 \text{Generators} & T_a & t_a & iT_a=t_a \\
\hline
 \text{Transformation of field  $\phi$} & &  & \\ 

 \text{ in a rep. of G} & \delta_{v}\phi=-v\phi & \delta_{\epsilon}\phi=i\epsilon\phi & iv=\epsilon; v_a=\epsilon_a\\
\hline
 \text{Gauge connection} & A'=A_a T^a & A=A_a t^a & iA'=A; A_a=A_a\\
\hline
 \text{Gauge connection variation} & \delta A'=dA'+[A',v] & \delta A=d\epsilon-i[A,\epsilon] &  \\
\hline
 \text{Gauge field-strength} & F'=F_a T^a & F=F_a t^a & iF'=F; F_a=F_a\\
\hline
 \text{Gauge field-strength} & F'=dA'+A'\wedge A' & F=dA-iA\wedge A &  \\
\hline
 \text{F variation} & \delta F'=[F',v] & \delta F=-i[F,\epsilon] &  \\
\hline
 \text{Covariant derivative} & D=d+A' & D=d-iA &  \\
\hline
\end{array}\)
\end{center}
\normalsize
\vspace{.5cm}

 If $\Delta$ is the anomalous phase variation and $I_{2r+2}$ the anomaly polynomial, the descent equations are given by:
\begin{center}
\begin{minipage}{5.0cm}
$\Delta=\delta\Gamma_M =-\int Q^1_{2r}$,\\
$I_{2r+2}=dI_{2r+1}$,\\
$\delta I_{2r+1}=dQ^1_{2r}$.
\end{minipage}
\end{center}\vspace{.5cm}

The following anomaly polynomials are used throughout the paper:
\begin{enumerate}
\item \textbf{Spin-1/2 fermion anomaly polynomial:}
\begin{equation}
I_{1/2}=(2\pi)\times [\widehat{A}(\mathcal{M}_d)]\times[ch(-iF)]\,,
\end{equation}
where
\begin{eqnarray}\label{DiracIndex}
\widehat{A}(\mathcal{M}_d)=1+\frac{1}{12\,(4\pi)^2}trR^2+\frac{1}{(4\pi)^4}\left[\frac{1}{360}trR^4+\frac{1}{288}(trR^2)^2\right]\nonumber\\
+\frac{1}{(4\pi)^6}\left[\frac{1}{5670}trR^6+\frac{1}{4320}trR^4 \, trR^2+\frac{1}{10368}(trR^2)^3\right]\nonumber\\
+\ldots
\end{eqnarray}
and
\begin{equation}
ch(-iF)=\sum_{k=0}\frac{1}{k!(2\pi)^k}TrF^k.
\end{equation}
\item \textbf{Gravitino (Spin-3/2) anomaly polynomial:}
\begin{eqnarray}
I^{d}_{3/2}&=& (2\pi)\times [\widehat{A}(\mathcal{M})][Tr(e^{\frac{iR}{2\pi}})-1]\times[ch(-iF)]\nn \\
&=&(2\pi)\times[ch(-iF)]\nonumber\\
&\times&\left[(d-1)+\frac{d-25}{12\,(4\pi)^2}trR^2+\frac{1}{(4\pi)^4}\left(\frac{d+239}{360}trR^4+\frac{d-49}{288}(trR^2)^2\right)\right.\nonumber\\
&+&\left.\frac{1}{(4\pi)^6}\left(\frac{d-505}{5670}trR^6+\frac{d+215}{4320}trR^4\,trR^2+\frac{d-73}{10368}(trR^2)^3\right)\right.\nonumber\\
&+&\ldots\Big]\,.
\end{eqnarray}
\item \textbf{Self-dual form:}
\begin{equation}
I_{\rm form}=(2\pi)\times [\widehat{L}(\mathcal{M}_d)]\times[-\frac{x}{4}]\,,
\end{equation}
where
\begin{eqnarray}
x=\begin{cases}
  1 \text{  if the base fermions are Weil or Majorana},\\
  1/2 \text{  if the base fermions are Majorana-Weil}
  \end{cases}
\end{eqnarray}
and
\begin{eqnarray}
\widehat{L}(\mathcal{M}_d)=1-\frac{1}{6\,(2\pi)^2}trR^2+\frac{1}{(2\pi)^4}\left(-\frac{7}{180}trR^4+\frac{1}{72}(trR^2)^2\right)\nonumber\\
+\frac{1}{(2\pi)^6}\left(-\frac{31}{2835}trR^6+\frac{7}{1080}trR^4\,trR^2-\frac{1}{1296}(trR^2)^3\right)\nonumber\\
+\ldots
\end{eqnarray}
\end{enumerate}

Finally, the Chern-Simons forms and descents are:\\
\begin{enumerate}
\item $TrF=dQ_1$, $Q_1=TrA$, $\delta Q_1=Trd\Sigma(x)$, $Q^1_2=Tr\Sigma(x)$.\\
\item $TrF^2=dQ_3$, $Q_3=Tr(A\wedge F-i\frac{1}{3}A^3)$, $\delta Q_3=Trd\Sigma(x)(dA)$, $Q^1_4=Tr\Sigma(x)(dA)$.\\
\item $TrF^3=dQ_5$, $Q_5=Tr(A\wedge F^2-\frac{1}{2}A^3F+\frac{1}{10}A^5)$, $\delta Q_5=Trd\Sigma(x)(dAdA-i\frac{1}{2}dA^3)$, $Q^1_6=Tr\Sigma(x)(dAdA-i\frac{1}{2}dA^3)$.
\end{enumerate}\vspace{.5cm}

\section{Trace formulae and branching rules}\label{Traces&BR}
In this Appendix, we record the conversion rules of the adjoint traces $Tr$ to fundamental traces $tr$, according to the standard convention of \cite{vanNieuwenhuizen:1989gc, Erler:1993zy}:

 \begin{subequations}\label{traces}
   \begin{align}
   TrF^2_{SO(N)}&= (N-2) ~trF^2_{SO(N)} \label{trace1}\,,\\
   TrF^2_{E_8} &= 30 ~trF^2_{E_8}\label{trace2}\,,\\
  TrF^4_{SO(N)}&= (N-8)~trF^4_{SO(N)}~ +~ 3~(trF^2_{SO(N)})^2 \label{trace3}\,,\\
  TrF^4_{E_8} &=\frac{1}{100}(TrF^2_{E_8})^2 = 9~(trF^2_{E_8})^2 \label{trace4}\,.
   \end{align}
      \end{subequations}
   As $E_8$ does not have a vector representation, it is standard to define its trace $trF^2_{E_8}$ in ``fundamental" by using that of the group $SO(32)$ so that they have uniform expressions in the Green-Schwarz term of both 10D Heterotic theories \cite{Green:1984sg}.

We also mention here the group theory branching rule of the adjoint representation, which is useful to interpret the various effective couplings arising from the amplitudes. For the breaking $\ett \to SO(16)^2 $ we have
\begin{equation}
\textbf{248 }\oplus \textbf{ 248}=\underbrace{\textbf{(120,1) $\oplus$ (1,120) }}_{\text{adjoint rep. of }SO(16)^2}\oplus \underbrace{\textbf{ (128,1) $\oplus$ (1,128)}}_{\text{spinor rep. of }SO(16)^2}.
\end{equation}
For the $\textbf{ (128,1) $\oplus$ (1,128)} $ representation, the trace-conversion formulae are
\begin{subequations}
\begin{align}
tr_{128}F_1^2 + tr_{128}F_2^2 &= 16 trF_1^2 +16 trF_2^2\,,\label{16trace1}\\
tr_{128}F_1^4 + tr_{128}F_2^4 &= 6 (trF_1^2)^2 + 6 (trF_2^2)^2 -8trF_1^4 -8trF_2^4\,.\label{16trace2}
\end{align}
\end{subequations}
For the breaking $SO(32) \to SO(16)^2 $ we have
\begin{equation}\label{so32BR}
\textbf{496}=\underbrace{\textbf{(120,1) $\oplus$ (1,120) }}_{\text{adjoint rep. of }SO(16)^2}\quad\oplus \underbrace{\textbf{ (16,16)}}_{\text{cospinor rep. of }SO(16)^2}.
\end{equation}
For the $\textbf{(16,16)}$ representation, the trace-conversion formulae are
\begin{subequations}\label{16trace}
\begin{align}
tr_{(16,16)}F^2&= 16 trF_1^2 +16 trF_2^2\,,\label{16trace3}\\
tr_{(16,16)}F^4 &= 16trF_1^4+16trF_2^4+6 trF_1^2trF_2^2\,.\label{16trace4}
\end{align}
\end{subequations}
For the breaking $E_8 \to SO(8)^2$ we have
\begin{align}
\textbf{248}=\underbrace{\textbf{(28,1) $\oplus$ (1,28) }}_{\text{adjoint rep. of }SO(8)^2}
&\oplus\underbrace{\textbf{(8,8)}}_{\text{bifundamental rep. of }SO(8)^2} \nonumber\\
& \oplus \underbrace{\textbf{(8,8)}'}_{\text{spinor rep. of }SO(8)^2}\oplus \underbrace{\textbf{(8,8)}''}_{\text{cospinor rep. of }SO(8)^2}.
\end{align} 
Thus the complete decomposition $E_8^{(1)} \rightarrow SO(8)_{(1)} \times SO(8)_{(2)}$ plus $E_8^{(2)} \rightarrow SO(8)_{(3)} \times SO(8)_{(4)}$ gives
\begin{align}
\label{e8decomp}
\textbf{248}\oplus \textbf{248} =& \textbf{(28,1,1,1)}\oplus \textbf{(1,28,1,1)}\oplus \textbf{(1,1,28,1)}\oplus \textbf{(1,1,1,28)}\\\nonumber
&\oplus \textbf{(8,8,1,1)}\oplus \textbf{(1,1,8,8)}\\\nonumber
& \oplus \textbf{(8,8,1,1)}'\oplus \textbf{(1,1,8,8)}'\\\nonumber
& \oplus \textbf{(8,8,1,1)}''\oplus \textbf{(1,1,8,8)}''.
\end{align}
The breaking $SO(32) \rightarrow SO(8)_{(1)} \times SO(8)_{(2)} \times SO(8)_{(3)} \times SO(8)_{(4)}$ gives instead:
\begin{align}
\label{32decomp}
\textbf{496} =& \textbf{(28,1,1,1)}\oplus \textbf{(1,28,1,1)}\oplus \textbf{(1,1,28,1)}\oplus \textbf{(1,1,1,28)}\nonumber\\
&\oplus \textbf{(8,8,1,1)}\oplus \textbf{(1,1,8,8)}\nonumber\\
& \oplus \textbf{(8,1,8,1)}\oplus \textbf{(1,8,1,8)}\nonumber\\
& \oplus \textbf{(1,8,8,1)}\oplus \textbf{(8,1,1,8)}.
\end{align}
From the decomposition \eqref{e8decomp} we see that $E_8^{(1)} \rightarrow SO(8)_{(1)} \times SO(8)_{(2)}$ plus $E_8^{(2)} \rightarrow SO(8)_{(3)} \times SO(8)_{(4)}$ has a preferred $trF_1^2 trF_2^2$ and $trF_3^2 trF_4^2$ mixing. T-duality exchanges the spinor and co-spinor representation with the bi-fundamental representations and this fact appears in the string 1-loop elliptic genus as orbifold shifts \cite{Kiritsis:2000zi}, which gives mixed couplings of the type $trF_1^2 trF_3^2$ and $trF_1^2 trF_4^2$  etc., even if one starts with the decomposition $E_8^{(1)} \rightarrow SO(8)_{(1)} \times SO(8)_{(2)}$ and $E_8^{(2)} \rightarrow SO(8)_{(3)} \times SO(8)_{(4)}$.

Finally, the trace-conversion formulae for the various representations involved in the breaking to $SO(8)^4$ are summarized as:
\begin{subequations}
\label{8trace}
\begin{align}
& tr_{(8,8)}F^2= 8trF_1^2 +8trF_2^2, \\
& tr_{(8,8)}F^4=8trF_1^4 +8trF_2^4+6trF_1^2trF_2^2,\\
& tr_{(8,8)'}F^2=tr_{(8,8)''}F^2= 8trF_1^2 +8trF_2^2,\\
& tr_{(8,8)'}F^4=tr_{(8,8)''}F^4 = 3(trF_1^2)^2+3(trF_2^2)^2 +6trF_1^2trF_2^2 -4trF_1^4 -4trF_2^4.
\end{align}
\end{subequations}


\section{Modular functions and decompactification limit}\label{sec:decomp}
In this Appendix, we gather the necessary definitions of the modular functions which have been used in the article:\\
 The Dedekind Eta function:
  \begin{equation}
\eta(\tau)=q^{1/24}\prod_{n=1}^{\infty}(1-q^n),
\end{equation}
satisfying \begin{equation}
\eta(-1/\tau)=\sqrt{-i\tau}\eta(\tau).
\end{equation}
The Leech j-function:
  \begin{equation}
  {j}= \frac{{E}^3_4}{{\eta}^{24}}=\frac{1}{{q}}+744+\cdots \, ,
  \end{equation}
  where $E_4$ is the 4-th Eisenstein series defined by
  \begin{equation}
  E_4=\frac{1}{2}\sum_{a=2}^4 \theta_a^8=1+240\sum_{n=1}^{\infty}\frac{n^3q^n}{1-q^n}.
  \end{equation} 
The following modular properties are used in fixing the $\sz$ anomalous phase variation in the anomaly counter-terms:
  \begin{eqnarray}
  log~ \eta(\tau+1)&=&log~ \eta(\tau)+i\frac{\pi}{12},\\
  log ~\eta(-\frac{1}{\tau})&=&log ~\eta(\tau)-i\frac{\pi}{4}+\frac{log ~\tau}{2}\\
  \frac{j(\tau+1)}{\bar{j}(\bar{\tau}+1)}&=&e^{-4i\pi}\frac{j(\tau)}{\bar{j}(\bar{\tau})},\\
\left(\frac{j(-1/\tau)}{\bar{j}(-1/\bar{\tau})}\right) ^{1/12}&=&-\left( \frac{j(\tau)}{\bar{j}(\bar{\tau})}\right)^{1/12}.
  \end{eqnarray}
  
\indent We now briefly discuss the  large volume and decompactification limits. The large volume limit in case of a $T^2$ compactification means taking the torus volume $V_{T^2} \rightarrow \infty$. However the complex structure $\U$  remains fixed.
We recall that the compact space-time torus is formed by compactifying the 8th and 9th space dimensions for which we have the following metric
\begin{equation}
\label{tormetric}
  G_{ij}=\left( \begin{smallmatrix} g_{88}& g_{89} \\ g_{89}& g_{99}\end{smallmatrix} \right)=\frac{V}{U_2}\left( \begin{smallmatrix} 1&U_1\\ U_1&|U|^2 \end{smallmatrix} \right).
\end{equation} 
  In the decompactification limit, we will take $V_{T^2} \rightarrow \infty$ and moreover impose orthonormality of the 8th and the 9th directions, i.e. 
\begin{equation}
\label{dlim1}
U_2 = \frac{V}{g_{88}}\rightarrow 1\,,\quad U_1=\frac{g_{89}}{g_{88}}\rightarrow 0\,.
\end{equation}
A useful summary of the q-expansion and the relevant limits of the different modular functions of $T$ and $U$ that have been used in expressions for the  higher-derivative couplings:
\begin{eqnarray}
\label{delim2}
log|\eta(T)|^2 &=& -\frac{\pi T_2}{6}-[\theta(T)+\bar{\theta}(\bar{T})],\\
\theta(\tau) &=& q+\frac{3 q^2}{2}+\cdots, \quad q=e^{2i\pi \tau},\\
\lim_{\tau\rightarrow i\infty}\theta(\tau) &=& 0, \quad \lim_{V\rightarrow \infty}log|\eta(T)|^2= -\frac{\pi T_2}{6},\\
log\frac{\eta(T)}{\bar{\eta}(\bar{T})}&=&\frac{i\pi T_1}{6}-[\theta(T)-\bar{\theta}(\bar{T})],\\
& &\lim_{V\rightarrow \infty}\left( log\frac{\eta(T)}{\bar{\eta}(\bar{T})} -\frac{i\pi T_1}{6}\right)=0\label{delim4} .
\end{eqnarray}
\begin{eqnarray}
\label{delim3}
U&=&U_1+iU_2 \rightarrow i,\quad q(U) \rightarrow  e^{-2\pi},\\
log|\eta(U)|^2 &\rightarrow & const,\\
log\frac{\eta(U)}{\bar{\eta}(\bar{U})}&=&\frac{i\pi U_1}{6}\rightarrow 0 \label{delim5}.
\end{eqnarray}



\end{document}